# Simple and Generic Simulator Algorithm for Inhomogeneous Random Spatial Deployment

## Mouhamed Abdulla and Yousef R. Shayan

**Abstract**—Due to technical intricacy, restricted resources, and high-cost for collecting empirical datasets, spatial modeling of wireless information networks via analytical means has been considered as a widely practiced mechanism for inference. As a result, diverse deployment models have been proposed for emulating the geometry of a network in order to explore its features. Although, these varied models are relevant in certain instances, but on the whole, such methods do not necessarily echo the actual inhomogeneous geometry of a network configuration over a particular deployment site. Therefore, we conceptualized a straightforward and flexible approach for random spatial inhomogeneity by proposing the area-specific deployment (ASD) algorithm, which takes into account the clustering tendency of users. In fact, the ASD method has the advantage of achieving a more realistic heterogeneous deployment based on limited planning inputs, while still preserving the stochastic character of users' position. We then applied this technique to different circumstances, and developed spatial-level network algorithms for controlled and uncontrolled cellular network deployments. Overall, the derived simulator tools will effectively and easily be useful for designers and deployment planners modeling a host of multi-coverage and multi-scale wireless network situations.

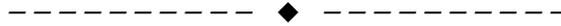

## 1 INTRODUCTION

MANY decades have passed since the original conception of the cellular network; however, despite the years and accumulated knowledge, there still remain numerous technical facets that have not been investigated thoroughly. The most notable among them is the paradigm of spatial random networks. In principle, within this context, it is desired to overlay nodes geometrical position onto the traditional fundamentals and understanding of mobile systems; where the broad motive is to analytically extract critical network-based observations.

### 1.1 Motivation

Evidently, emulation is a powerful approach that assists applied scientists in better understanding the system under investigation. Indeed, once adequately conceptualized, this technique develops into an indispensible analysis mechanism because the method for reengineering and modeling the network architecture can be shown to be:

- Cost and time efficient.
- Adjustable by simple parameter modifications.
- Informative in studying a complex platform.

As a result, having the capability to duplicate via models the footprint of real-world wireless networks, and then draw important fundamentals from these characterizations, is important for effective network design and planning during both pre- and post-deployments.

### 1.2 Related Work

Indeed, some fairly acceptable conjectures have been adopted in literature in order to alleviate the burden of spatial emulation. Notably, the random homogeneous model is a conceivable assertion for stochastical inference, particularly when users' spatial pattern is lacking [1], [2], [3], [4], [5]. However, due to their social fabric, mobile carrying end-users tend to gather with a higher likelihood in some preferred locations as opposed to an equalized arrangement; thus implying the inevitability of heterogeneous distributions.

As a counter reaction for this need, various inhomogeneous spatial deployment models have been suggested. For instance, the principle of thinning can be applied as one possible approach, where an inhomogeneous spatial distribution is synthetically realized by deleting nodes from a uniformly deployed pattern [6], [7]. Another technique enables heterogeneity through different adaptation of edge or center-focused deployments by theoretically adjusting the spatial models through a simple tunable variable [8], [9]. As an additional alternative, the Gaussian geometry is distinctively an interesting heterogeneous model because both the geographical spread and the intensity of terminals position are flexible [10]. Indeed, the dual purpose of this network deployment model can be controlled by its standard deviation. Therefore, this random structure can be utilized for emulating various multi-pattern user-carried devices in a cellular architecture [5], [9], [11], [12], [13], [14], [15].

Although practical for preliminary analysis, but by and large, these spatial deployment models and the like will not necessarily generate reliable mapping of the network footprint. Therefore, they could inaccurately reflect important technical issues of relevance to network planning and design.

_____________________

- *The authors are with the Department of Electrical and Computer Engineering, Concordia University, Montréal, Québec, Canada, 1515 Ste. Catherine W., H3G-2W1. E-mail: {m_abdull, yshayan}@ece.concordia.ca.*

### 1.3 Objective

As a consequence of the above, it becomes imperative to find new practices for inhomogeneous random deployment. Intrigued by this challenge, in this paper, we intend to contrive spatial mechanisms for constructing adaptable networks that can realistically map users' trends while still preserving the random character of deployments. Moreover, we want these bona fide heterogeneous models to require limited *a priori* input parameters from designers so as to ensure their ease of configuration for an array of network planning projects.

To this end, while bearing in mind that reflective network emulation is usually very complex to realize, we nonetheless aim to tackle this deployment objective by probing the essential underpinning of nodal clustering. As a matter of fact, users' spatial structures are mainly shaped and characterized by natural and manmade topographical land-cover features and environments. Thus, our solution to this inhomogeneous undertaking would be to conceptualize ASD, which is a random deployment approach such that users tendency to cluster based on terrain limitations is exclusively taken into account.

After formulating the corresponding algorithms, we then intend to demonstrate various random network realizations generated from Monte Carlo (MC) simulations. Altogether, the obtained results are expected to deliver a practical toolkit that will be instrumental in researching the facets of radio networks related to connectivity and service quality.

### 1.4 Organization

The rest of this paper is organized as follows. In Section 2, we will explicitly formulate and analyze the geometrical characteristics of a flexibly versatile random network model. Then, in Section 3, we will explain the ASD algorithm for the purpose of emulating spatial inhomogeneity. After, in Section 4, we will utilize this proposed algorithm to develop a heterogeneous mechanism for controlled random deployment. Next, in Section 5, we will also conceive a technique for automatically generating an arbitrary geometrical structure with least amount of inputs. Afterward, in Section 6, we will outline a general synopsis of the developed simulator models. Finally, Section 7 will conclude the paper.

## 2 CHARACTERISTICS OF A GEOMETRICALLY VERSATILE NETWORK MODEL

### 2.1 Simple Model for the Radiation Coverage

For analytical convenience, the depicted isotropic antenna shown in Fig. 1, which has a perfect spherical shape, has been conceived to idealistically model the EM radiation of a toroidal-like omni-directional emitter. Irrespective of whether the model pattern is isotropic or omni-directional, when projected on a Euclidian plane, the extent of the EM propagation will result in a perfect circular contour with base-station (BS) located at its centroid.

Realistically, the BS radiation shape is in fact irregular in format due to external agents such as: channel losses caused by terrain features, manmade obstacles, and at-mospheric attenuation [16]. Fig. 2 visually depicts the actual and ideal radiation profile of a tower station for centralized connectivity. In principle, the cellular adjustment from the actual to the ideal is performed in order to straightforwardly facilitate various cellular-based technical analyses, including network deployment.

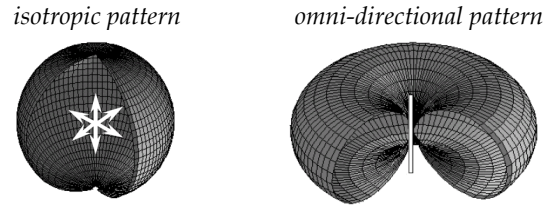

Fig. 1. All-direction antenna radiation models (3D).

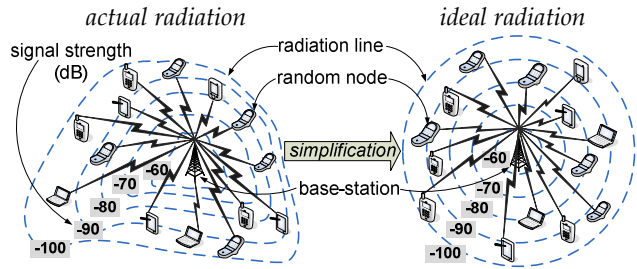

Fig. 2. Impact of channel features on EM radiation pattern (2D).

### 2.2 Exact Random Spatial Deployment

For various mobile communication purposes the circular random network model has been presumed, among others, in [1], [2], [3], [4], [5]. Despite the availability of this spatial model, in its current state it does not offer any deployment versatility in dealing with sectored layers and edge related aspects. Clearly, the needed geometrical adaptability could be created synthetically through heuristic means. However, besides being an inefficient generation approach, such workarounds alters the wanted randomness; thus defeating the main principle of stochastic networks [17]. As a consequence, we will in this subsection derive the exact and appropriate expressions needed for versatile random nodal deployment.

To begin, instead of making the cell shape represents the BS radiation coverage, we rather make it correspond to the surface area of some terrain. For the sake of the argument, let us assume that the surface region of interest has circular ring sector geometry. And, for the simplest and possibly most intuitive case for spatial deployment, we may postulate that nodes are uniformly distributed within this geographical strip. As a result, the joint spatial PDF for nodes 2D position will have the form depicted in Fig. 3, where the inner and outer cellular radii of the ring sector are identified by $\exists L_1, L_2 \in \mathbb{R}_+^2 : 0 \leq L_1 < L_2$, and the angular limits are given by $\exists \alpha_1, \alpha_2 \in \mathbb{R}_+^2 : 0 \leq \alpha_1 < \alpha_2 \leq 2\pi$. As for the deployment region, it can be assessed over the surface domain by simply integrating an infinitesimal area element, i.e.: $dA = dx \cdot dy = r \cdot dr \cdot d\theta$. Pursuing this task, produces the result in (1) such that $A_{RS} \in \mathbb{R}_+^*$ is the corresponding deployment area of the network cluster for the ring sector, and $D_{RS}$ is the support domain in Cartesian format. The spatial density can then be formulated

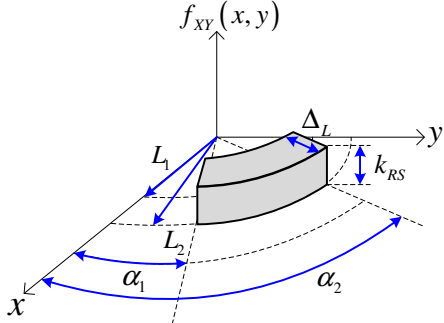

Fig. 3. Spatial density over a ring sector in Cartesian coordinates.

$$A_{RS} = \iint_{(x,y) \in D_{RS} \subset \mathbb{R}^2} dA = \left( L_2^2 - L_1^2 \right) \left( \alpha_2 - \alpha_1 \right) / 2 \quad (1)$$

by its reciprocal, namely: $f_{XY}(x,y) = 1/A_{RS}$.

For generation purposes, the next step demands that we determine the marginal density along each axis. If we continue with rectangular coordinates, the analysis will become longer and more complicated to solve. Given the character of the network cluster being modeled, it is evident that the best stochastic transformation ought to depend on the polar system:

$$f_{R\theta}(r,\theta) = f_{XY}(x,y)\Big|_{\substack{x=r\cos\theta \\ y=r\sin\theta}} \cdot \left| J(r,\theta) \right|$$
$$= 2 \cdot r \big/ \left( L_2^2 - L_1^2 \right) \left( \alpha_2 - \alpha_1 \right) \quad (2)$$

where its support surface is described by:

$$D_{RS}^P = \left\{ \begin{array}{l|l} (r,\theta) \in \mathbb{R}_+^2 & 0 \le L_1 \le r \le L_2 \\ (L_1, L_2, \alpha_1, \alpha_2) \in \mathbb{R}_+^4 & 0 \le \alpha_1 \le \theta \le \alpha_2 \le 2\pi \end{array} \right\} \quad (3)$$

The portrayal of the modified density function is accordingly depicted in Fig. 4.

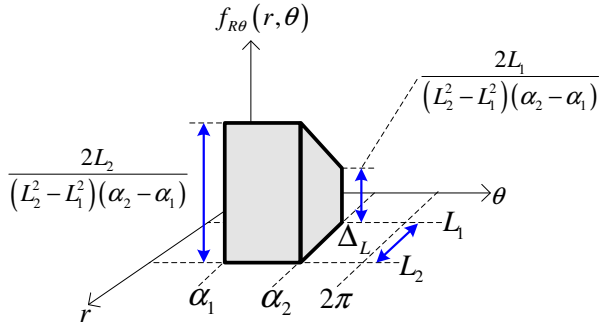

Fig. 4. Spatial density over a ring sector in polar coordinates.

Using the results of (2) and (3), we could at present express the marginal probability densities for the radial and angular components:

$$f_R(r) = \int_{(r,\theta) \in D_{RS}^P} f_{R\theta}(r,\theta) \, d\theta = 2r \big/ \left( L_2^2 - L_1^2 \right) \quad (4)$$

$$f_\theta(\theta) = \int_{(r,\theta) \in D_{RS}^P} f_{R\theta}(r,\theta) \, dr = \mathcal{U}_\theta(\alpha_1, \alpha_2) \quad (5)$$

From (4) and (5), we can readily show that random variables (RVs) $R$ and $\theta$ are actually independent because: $f_{R\theta}(r,\theta) = f_R(r) \cdot f_\theta(\theta)$. Pursuing this further, the associated radial CDF of (4) can then be computed by:

$$F_R(r) = \Pr(R \le r) = \int_{\tilde{r}=-\infty}^{r} f_R(\tilde{r}) \, d\tilde{r} = \left( r^2 - L_1^2 \right) \big/ \left( L_2^2 - L_1^2 \right) \quad (6)$$

If we set the CDF of (6) to an arbitrary sample occurrence $\hat{u}$ generated from a standard uniform distribution, then the related inverse CDF should enable efficient emulation of instances, i.e.: $\hat{r} = \left\{ (F_R)^{-1} (\hat{u} \sim \mathcal{U}(0,1)) \right\} \sim f_R(r)$. After solving this expression, we notice that the radial samples will be generated by:

$$\hat{r} = \sqrt{L_1^2 + \hat{u}\left( L_2^2 - L_1^2 \right)} : \left\{ \exists \hat{u} \in \mathbb{R}_+^* : 0 < \hat{u} < 1 \right\} \mapsto (L_1, L_2) \quad (7)$$

As for the angular component, its samples are produced by:

$$\hat{\theta} = \left\{ \alpha_1 + \hat{v}(\alpha_2 - \alpha_1) \right\} \sim f_\theta(\theta) \quad (8)$$

where $\hat{u}$ and $\hat{v}$ in (7) and (8) are uncorrelated i.i.d. samples.

To verify the generation accuracy of the radial density, we performed in Fig. 5 a set of random simulations. Specifically, the outer radius of the network ring was fixed, and the inner radius varied for different values. For each network case, the simulation was performed based on $n_S = 10,000$ samples with a histogram resolution of $n_B = 100$. As evident, the radial PDF based on theoretical analysis and MC measures are in agreement.

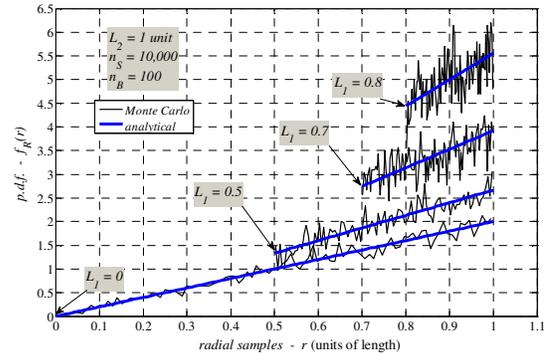

Fig. 5. Radial density of the deployment via MC simulations.

Furthermore, to demonstrate the flexibility and the generic nature of the network model just derived, we obtained through random MC simulation the results of Fig. 6 for different parameter values and nodal densities assed by $\rho_\lambda$. The six unique random network examples of the figure were realized by two type of inputs; namely the geometrical attributes of the random cluster: $L_1, L_2, \alpha_1, \alpha_2$; and the scale of the network: $n_S$. As visually manifest, the 2D deployments and the spatial densities match the anticipated footprint of the network.

## 2.3 Analysis of the Spatial Density

In this subsection, we intend to further probe the estimation of spatial density between theoretically predicted formulation and randomly simulated results. Precisely, once the random 2D deployment is realized, we then consider these arbitrary geometrical samples in order to represent a bivariate histogram that approximates the Euclidian distribution of the deployment. For this purpose, as illustrated in Fig. 7, we consider the approximation of a general probability function by its histogram equivalent. To analytically characterize this estimation, the bivariate fundamental histogram bin positioned at the origin of a Cartesian coordinate system is assigned to: $\delta_B(x,y) = \mathbf{1}(|x| \leq \Delta x_B/2; \ |y| \leq \Delta y_B/2)$, such that $\mathbf{1}_A(x,y)$ is a 2D indicator function over $(x,y) \in A \subseteq \mathbb{R}^2$, and $(\Delta x_B, \Delta y_B) \in \mathbb{R}^2_{+,*}$ are the dimensions of each histogram bin. In fact, these dimensions can be computed by: $\Delta x_B = (x_H - x_L)/n_{B-X}$ and $\Delta y_B = (y_H - y_L)/n_{B-Y}$, where $[x_L, x_H] \times [y_L, y_H]$ identifies the deployment surface of the network, and $(n_{B-X}, n_{B-Y})$ represents the resolution of the bivariate histogram. Furthermore, the number of occurrence for the $(i,j)$-th bin is defined by:

$$h_{(i,j)} : \left\{ (i,j) \in \mathbb{N}^2_* \mid 1 \leq i \leq n_{B-X}; \ 1 \leq j \leq n_{B-Y} \right\} \mapsto \mathbb{N} \quad (9)$$

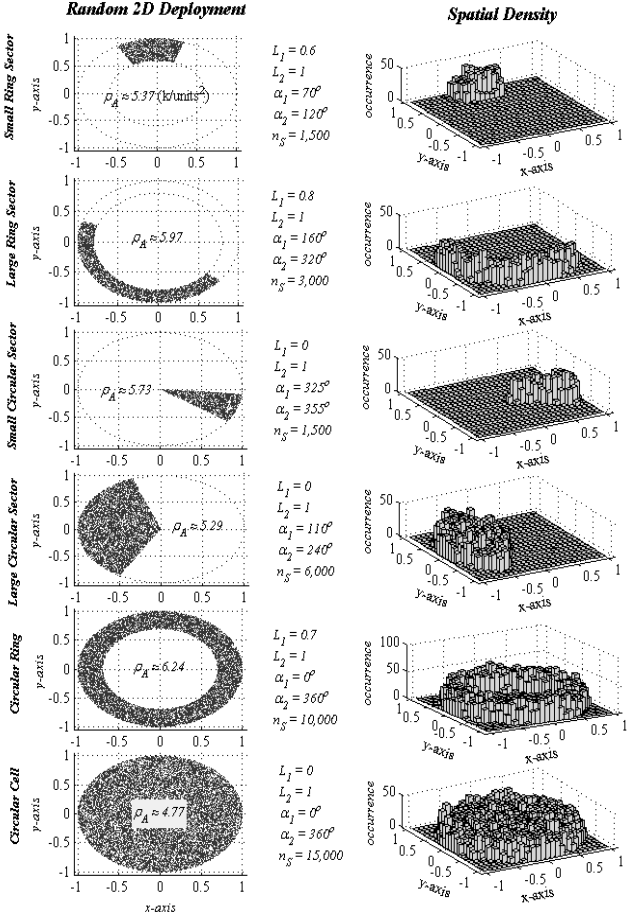

Fig. 6. Deployment versatility of random network models.

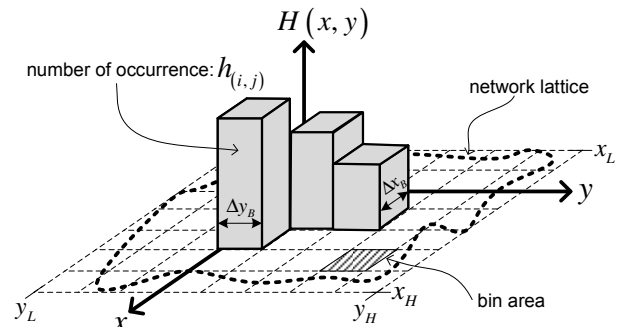

Fig. 7. A general bivariate histogram realization.

Using the above definitions and assignments, we could therefore express the overall spatial density histogram $H(x,y) : \mathbb{R}^2 \mapsto \mathbb{N}$ for a randomly deployed network by:

$$H(x,y) = \sum_{j=1}^{n_{B-Y}} \sum_{i=1}^{n_{B-X}} h_{(i,j)} \cdot \delta_B(x - x_i, y - y_j) \quad (10)$$

having center positions $(x_i, y_j) \in \mathbb{R}^2$ parameterized by:

$$x_i = x_L + (i - 1/2)\Delta x_B \quad i = 1, 2, \cdots, n_{B-X} \quad (11)$$

$$y_j = y_L + (j - 1/2)\Delta y_B \quad j = 1, 2, \cdots, n_{B-Y} \quad (12)$$

At this point, we could analytically obtain the average bivariate histogram density by $h_{XY}^{\text{analytical}} = \langle h_{XY} \rangle = \rho_A \cdot A_{bin}$, such that $\rho_A \in \mathbb{R}^*_+$ is the number density of the spatial network, and $A_{bin} \in \mathbb{R}^*_+$ represents the surface area of the bivariate bin. Consequently, this could be rewritten in general terms by: $h_{XY}^{\text{analytical}} = n_S \cdot A_{bin}/A_N = n_S \cdot \Delta x_B \cdot \Delta y_B/A_N$, where $n_S \in \mathbb{N}^*$ is the number of randomly generated samples, and $A_N \in \mathbb{R}^*_+$ is the surface area of the lattice.

Specifically, if we want to estimate the histogram density of the versatile network model, we could consider the spatial footprint depicted in Fig. 8. As illustrated, the histogram grid is based on equally-spaced bin regions of $\Delta_B \in \mathbb{R}^*_+$ dimensions. For precisions purposes, it is worth noting that although the figure portrays a $10 \times 10$ grid, in our generic derivation we will assume a 2D resolution of $n_B^2$. The dimension of the bin area is therefore obtained by: $\Delta_B = \Delta x_B = \Delta y_B = 2L_2/n_B$. We thus find that:

$$h_{XY}^{\text{analytical}} = n_S \Delta_B^2 / A_{RS} = 8n_S / n_B^2 (\alpha_2 - \alpha_1)\left\{ 1 - (L_1/L_2)^2 \right\} \quad (13)$$

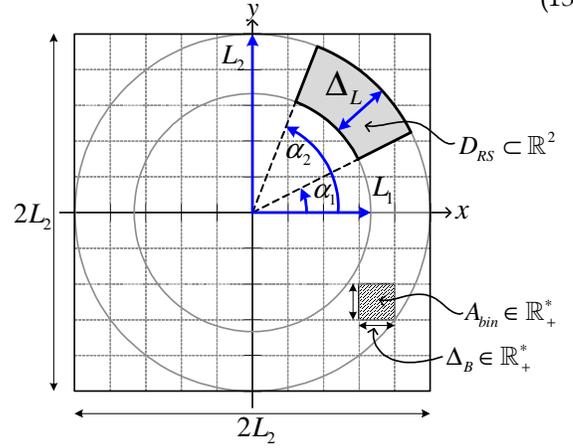

Fig. 8. Footprint of a versatile network model for density estimation.

The histogram density may also be computed from experimental MC data points through its empirical mean, which is defined in its general format by:

$$h_{XY}^{\text{simulation}} = \langle \hat{h}_{XY} \rangle = \frac{1}{n_{XY}} \sum_{j=1}^{n_{B-Y}} \sum_{i=1}^{n_{B-X}} h_{(i,j)} \quad (14)$$

where $n_{XY} \in \mathbb{N}^*$ is the amount of bins over the deployment surface having a nonzero occurrence number. To be accurate, this value is in fact bounded by: $0 < n_{XY} \leq n_B \cdot n_{B-Y}$.

In order to verify the statistical metrics developed above, we performed a number of MC simulations for various permutations of the randomly modeled network. Essentially, we considered the same network geometries as those described in Fig. 6; except, random instances of the position samples were augmented in order to produce a reliable approximation of the spatial density. As for the estimation step, the quantity of subdivisions along the $x$ and $y$ axes were equal, and set to $n_B = 500$ for all cases. Table 1 presents a contrast of the spatial density estimation between theoretical prediction and simulated data. It should be clear from the table that although both bivariate histogram density measures of (13) and (14) have units of no. per bin area, they will not be integer values, rather each will be in $\mathbb{R}_+^*$ because they represent average quantities. The percentage error of the spatial density among analysis and simulation were quantified by:

$$\varepsilon_A = \left\{ \left| h_{XY}^{\text{simulation}} - h_{XY}^{\text{analytical}} \right| \middle/ h_{XY}^{\text{analytical}} \right\} \times 100 \qquad (15)$$

Given the slight value of the error, we can conclude the validity of the formulated statistical estimation analysis.

TABLE 1
CONTRASTING SPATIAL DENSITY ESTIMATION

| random network models | $A_{RS}$ (units$^2$) | $n_S$ (no.) | $\rho_A$ (k/units)$^2$ | $n_B$ (no.) | $h_{XY}^{\text{analytical}}$ (no./bin area) | $h_{XY}^{\text{simulation}}$ (no./bin area) | $\varepsilon_A$ (%) |
|---|---|---|---|---|---|---|---|
| small ring sector | 0.2793 | $10^6$ | 3.5810 | 500 | 57.2958 | 56.4602 | 1.46 |
| large ring sector | 0.5027 | $10^6$ | 1.9894 | 500 | 31.8310 | 31.2297 | 1.89 |
| small circular sector | 0.2618 | $10^6$ | 3.8197 | 500 | 61.1155 | 59.9434 | 1.92 |
| large circular sector | 1.1345 | $10^6$ | 0.8815 | 500 | 14.1036 | 14.0280 | 0.54 |
| circular ring | 1.6022 | $10^6$ | 0.6241 | 500 | 9.9862 | 9.9137 | 0.73 |
| circular cell | 3.1416 | $10^7$ | 3.1831 | 500 | 50.9296 | 50.7354 | 0.38 |

Overall, in this section, we demonstrated and analyzed the approach for spatial flexibility in random deployments by deriving exact and generic stochastic expressions based on efficient random generation. As it will be shown in subsequent sections, the described geometrical model will serve as a fundamental steppingstone for developing controlled and uncontrolled heterogeneous network algorithms.

## 3 AREA-SPECIFIC DEPLOYMENT STRATEGY FOR SPATIAL INHOMOGENEITY

### 3.1 General Principle

At present, we need to construct a randomly tunable algorithm that takes into consideration the fundamental ingredients of spatial deployment. From a visceral observation, it becomes natural to give precise attention to following criteria [10]:

1. Geography of the Network: This constitutes the general location and setting of the network. Namely, is the spatial emulation intended for a rural, or rather a built-up urban region?

2. Topography of the Network: This part looks into the details of the terrain and its distinctive landforms and features.

3. Demography of the Network: Here, the scale and distribution of users is important. Namely, is the network densely or sparsely populated, and how does this composition change with time?

On the whole, it is desired to conceive an easily controlled and configured algorithm with least amount of inputs while overlaying the above three aspects in order to closely reflect the specifications and limitations of a particular terrain site. These are all diametrically opposing requirements, and so reconciling them simultaneously is rather difficult to solve. Despite being quite involved, it is still possible to undertake this objective by contriving a framework that adheres to the notion of *divide and conquer*. That is, spatial deployment can be tackled by gradually breaking down this challenge into smaller algorithmically solvable parts, and then synthesizing the results.

In particular, this is done by proposing a superposition-based algorithm which we refer to as area-specific deployment (ASD). As shown in the descriptive example of Fig. 9, the general abstraction of the ASD approach can be described gradually in a systematic manner.

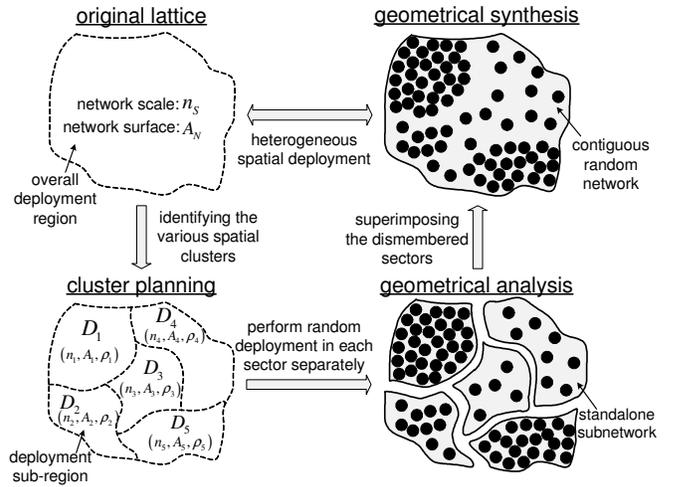

Fig. 9. Characterizing ASD by a descriptive network model.

Essentially, for a particular project site, the deployment designer will identify various likely clusters such that the overall network scale and surface is selectively split among these smaller sub-regions. As a matter of fact, each of the compiled non-overlapping clusters is uniquely specified by its: lattice shape or support domain $D_i \subseteq \mathbb{R}^2$; surface area $A_i \in \mathbb{R}_+^*$; quantity of randomly positioned nodes $n_i \in \mathbb{N}^*$; and corresponding number density $\rho_i \in \mathbb{R}_+^*$. Moreover, the areal size and nodal volume of the original lattice are respectively given by: $A_N = \sum_{i=1}^{n_{\text{sec-total}}} A_i$ and $n_S = \sum_{i=1}^{n_{\text{sec-total}}} n_i$, such that $n_{\text{sec-total}} \in \mathbb{N}^*$ is the overall amount of sectors.

Once the planning of the network footmark is set, we then focus on the sub-regions in a standalone way so as to stochastically generate the desired volume of Euclidian positions. Precisely, random uniform deployment is presumed over the specified sectors as expressed by the particular spatial density function of the sub-regions, i.e.:

$$f_{XY}(x, y) = 1/A_i \cdot \mathbf{1}_{D_i \subseteq \mathbb{R}^2}(x, y) \quad i = 1, 2, \cdots, n_{\text{sec-total}} \qquad (16)$$

However, since the clusters have different lattice shapes; probabilistic analysis has to carefully be drafted for random generation in a specific location, with a particular geometrical contour, coverage size, and nodal scale.

Following the analysis step, the various subnetworks are then reassembled in a puzzle-like format. Thus, as a consequence of network synthesis, heterogeneous spatial distribution emerges over the deployment field.

Evidently, the ASD technique gives the necessary leverage to designers to tailor and plan the spatial architecture when *a priori* knowledge of the network is asserted while still preserving the geometrical randomness of the users. Such attributes will hence ensure greater emulation flexibility and attain spatial heterogeneity so as to evaluate a host of network-based QoS factors.

### 3.2 Spatial Formulation for the ASD Algorithm

At present, we aim to derive a PDF estimation expression tailored specifically for approximating the spatial density of an inhomogeneous random network deployment realized via the proposed ASD algorithm. For this purpose, we define the histogram:

$$H_{ASD}(x, y) = \sum_{j=1}^{n_{B-Y}} \sum_{i=1}^{n_{B-X}} h_{(i,j)}^{ASD} \cdot \delta_B(x - x_i, y - y_j) \quad (17)$$

As it can be observed, this expression is similar to the notation of (10) except that here we consider multiple deployment regions. As a consequence, $h_{(i,j)}^{ASD} \in \mathbb{N}$ will equal the aggregate of the multi-density sectors:

$$h_{(i,j)}^{ASD} = h_{(i,j)}^{(1)} + h_{(i,j)}^{(2)} + \cdots + h_{(i,j)}^{(n_{sec-total})} = \sum_{k=1}^{n_{sec-total}} h_{(i,j)}^{(k)} \quad (18)$$

And for this generalized case, the overall deployment surface $[x_L, x_H] \times [y_L, y_H]$ is obtained by:

$$x_L = \min_{k=1,2,\cdots,n_{sec-total}} \left\{ x_L^{(k)} \right\}; \quad x_H = \max_{k=1,2,\cdots,n_{sec-total}} \left\{ x_H^{(k)} \right\} \quad (19)$$

$$y_L = \min_{k=1,2,\cdots,n_{sec-total}} \left\{ y_L^{(k)} \right\}; \quad y_H = \max_{k=1,2,\cdots,n_{sec-total}} \left\{ y_H^{(k)} \right\} \quad (20)$$

To determine the density of (17), we need to go over the fundamentals of stochastic theory. Indeed, the probability of some arbitrary event $A$ obtained for a bivariate PDF will be equal to:

$$\Pr(A \to (x, y) \in D) = \iint_{(x,y) \in D} f_{XY}(x, y) \, dx \, dy \quad (21)$$

This expression can be approximated by examining the left and right hand sides of (21) separately and then equating them together, namely:

$$\Pr(A) \approx H_{ASD}(x, y) / n_S \approx \tilde{f}_{XY}(x, y) \cdot \Delta x_B \cdot \Delta y_B \quad (22)$$

such that $\tilde{f}_{XY}(x, y)$ is the numerical PDF estimation for spatial inhomogeneous deployment. If we isolate for the density function, we obtain the final result as follows:

$$f_{XY}(x, y) = \lim_{n_S \to \infty} \lim_{\substack{n_{B-X} \to \infty \\ \Delta x_B \to 0}} \lim_{\substack{n_{B-Y} \to \infty \\ \Delta y_B \to 0}} \tilde{f}_{XY}(x, y) \quad (23)$$

where:

$$\tilde{f}_{XY}(x, y) = \sum_{j=1}^{n_{B-Y}} \sum_{i=1}^{n_{B-X}} \sum_{k=1}^{n_{sec-total}} \frac{h_{(i,j)}^{(k)} \delta_B(x - x_i, y - y_j)}{n_S \cdot \Delta x_B \cdot \Delta y_B} \quad (24)$$

As noted by the limits in (23), the spatial density estimation can be improved by augmenting the quantity of MC samples $n_S$. Also, increasing the histogram resolution through the number of bars along each axis is expected to ameliorate the numerical computation of the 2D density function. However, rising $n_S$ and $n_B$ (assuming $n_{B-X} = n_{B-Y}$) simultaneously by a certain level does not necessarily improve the result. This is in fact the case because by (24) these elements oppositely impact the spatial density function, i.e.:

$$\tilde{f}_{XY}(x, y) \propto 1/n_S \Delta x_B \Delta y_B \propto n_{B-X} n_{B-Y} / n_S \propto n_B^2 / n_S \quad (25)$$

Therefore, a better understanding of the joint relationship between these estimation factors is needed in order to fine-tune the approximation process.

Nonetheless, we should emphasize that the tractable explanation of (23) and (24) offer an analytical notation for estimating users' inhomogeneous geometrical trend over a geographical service area. In fact, the formulated result is tailored specifically for approximating the spatial density function of an ASD-based heterogeneous network. Indeed, the result is exclusively applicative and valid for the controlled and uncontrolled inhomogeneous deployment algorithms anticipated in Sections 4 and 5.

## 4 CONTROLLED ALGORITHM FOR RANDOM DEPLOYMENT

### 4.1 Formulation of the Network Model

In the previous section, we provided a high-level view for attaining inhomogeneity. In this part of the paper, we will apply the proposed ASD method in order to conceive an approach for generating spatial heterogeneity.

As a visual aid in deriving the non-uniform algorithm, we consider the canonical network model of Fig. 10. From the display, it should be evident that the approach for partitioning the cell is in part inspired by the various layer formations apparent in the cross-section of an onion. Clearly, there are no sectors in the onion-layer arrangements; yet to add another level of deployment versatility to the conceptualized spatial model, we enable the possibility of incorporating sector strips in each layer of the network plan. This modification will in essence augment and enhance the inhomogeneous capability of the model.

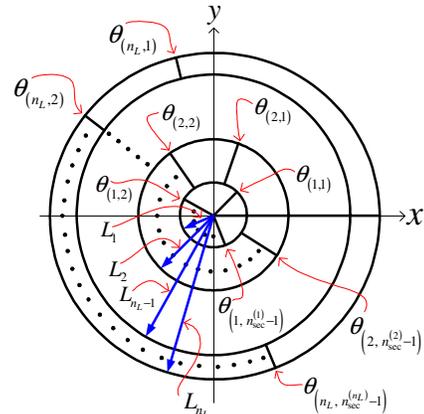

Fig. 10. Modeling the network plan for heterogeneous deployment.

In the above model, we recognize that the circular cell is split into $n_L \in \mathbb{N}^*$ layers. And, each layer contains

$n_{\text{sec}}^{(i)} \in \mathbb{N}^*$ sectors, such that $i = 1, 2, \cdots, n_L$. Therefore, the total number of sectors in the inhomogeneous model of Fig. 10 can be computed by: $n_{\text{sec-total}} = \sum_{i=1}^{n_L} n_{\text{sec}}^{(i)}$. In principle, the more terrain strips we consider during the planning stage of a particular project site, the more network clusters will be resulted, and thus $n_{\text{sec-total}}$ will raise. And as the total number of sectors with varying densities increase, it will consequently impact the geometrical inhomogeneity level of wireless nodes. In other words, the size of $n_{\text{sec-total}}$ is an indicator for the details and precision of the deployment plan, which is in fact left to the discretion of the network architect.

Furthermore, each cluster sector is indeed bounded within two radii and two angular limits. The layers radii for heterogeneous random deployment are collectively contained by the $\mathbf{R} \in \mathbb{R}_{+,*}^{n_L}$ vector, which is specified by:

$$\underset{n_L \times 1}{\mathbf{R}} = \begin{bmatrix} r_1 & r_2 & \cdots & r_{n_L} \end{bmatrix}^T = \begin{bmatrix} r_i \in \mathbb{R}_+^* \end{bmatrix}_{i=1,2,\cdots,n_L} \quad (26)$$

such that $r_{i-1} < r_i : i = 2, 3, \cdots, n_L$.

As for the sectors angular information, they are identified by their higher values within the $\underset{n_L \times (\gamma-1)}{\mathbf{\Theta}} \in \mathbb{R}_+^{n_L \times (\gamma-1)}$ matrix:

$$\underset{n_L \times (\gamma-1)}{\mathbf{\Theta}} = \begin{bmatrix} \theta_{(1,1)} & \theta_{(1,2)} & \cdots & \theta_{\left(1, n_{\text{sec}}^{(1)}-1\right)} & 0 & \cdots & 0 \\ \theta_{(2,1)} & \theta_{(2,2)} & \cdots & \theta_{\left(2, n_{\text{sec}}^{(2)}-1\right)} & 0 & \cdots & 0 \\ \vdots & \vdots & \vdots & \vdots & \vdots & \vdots & \vdots \\ \theta_{(k_L,1)} & \theta_{(k_L,2)} & \cdots & \cdots & \cdots & \theta_{\left(k_L, n_{\text{sec}}^{(k_L)}-1\right)} = \theta_{(k_L,\ \gamma-1)} \\ \vdots & \vdots & \vdots & \vdots & \vdots & \vdots & \vdots \\ \theta_{(n_L,1)} & \theta_{(n_L,2)} & \cdots & \theta_{\left(n_L, n_{\text{sec}}^{(n_L)}-1\right)} & 0 & \cdots & 0 \end{bmatrix}$$
$$= \begin{bmatrix} \theta_{(i,j)} \in \mathbb{R}_+^* \end{bmatrix}_{\substack{j=1,2,\cdots,n_L \\ j=1,2,\cdots,n_{\text{sec}}^{(i)}-1}}$$
$$(27)$$

However, since the last sector of any layer is always set to $2\pi$, then there is no need to enter this reoccurring measure in the matrix. In fact, the various angular values for each layer are bordered by:

$$0 < \theta_{(i,1)} < \theta_{(i,2)} < \cdots < \theta_{\left(i,\ n_{\text{sec}}^{(i)}-1\right)} < 2\pi \quad i = 1, 2, \cdots, n_L \quad (28)$$

Also, in (27) the $\gamma \in \mathbb{N}^*$ represents the largest number of sectors in a particular network layer, which is quantified by:

$$\gamma \triangleq \max_{i=1,2,\cdots,n_L} \left\{ n_{\text{sec}}^{(i)} \right\} = n_{\text{sec}}^{(k_L)} \quad (29)$$

such that $\left\{ k_L \in \mathbb{N}^* \,\middle|\, 1 \le k_L \le n_L \right\}$ is the network layer that has the greatest amount of sectors. It is worth noting that this value is not necessarily unique because there might be multiple layers that have the similar maximum number of sectors.

Pursuing this further, the $\mathbf{N} \in \mathbb{N}^{n_L \times \gamma}$ matrix of (30) holds the amount of randomly positioned nodes deployed in each sector. This means that the spatial topology is tunable by simply modifying the quantity of nodes in the cluster strips of the network plan.

For the convenience of manipulations, the radial, angular and nodal entries respectively expressed in (26), (27), and (30) can be assembled together by the network plan

$$\underset{n_L \times \gamma}{\mathbf{N}} = \begin{bmatrix} n_{(1,1)} & n_{(1,2)} & \cdots & n_{\left(1, n_{\text{sec}}^{(1)}\right)} & 0 & \cdots & 0 \\ n_{(2,1)} & n_{(2,2)} & \cdots & n_{\left(2, n_{\text{sec}}^{(2)}\right)} & 0 & \cdots & 0 \\ \vdots & \vdots & \vdots & \vdots & \vdots & \vdots & \vdots \\ n_{(k_L,1)} & n_{(k_L,2)} & \cdots & \cdots & \cdots & n_{\left(k_L, n_{\text{sec}}^{(k_L)}\right)} = n_{(k_L,\gamma)} \\ \vdots & \vdots & \vdots & \vdots & \vdots & \vdots & \vdots \\ n_{(n_L,1)} & n_{(n_L,2)} & \cdots & n_{\left(n_L, n_{\text{sec}}^{(n_L)}\right)} & 0 & \cdots & 0 \end{bmatrix}$$
$$= \begin{bmatrix} n_{(i,j)} \in \mathbb{N}^* \end{bmatrix}_{\substack{i=1,2,\cdots,n_L \\ j=1,2,\cdots,n_{\text{sec}}^{(i)}}}$$
$$(30)$$

matrix $\mathbf{P} \in \mathbb{R}_+^{n_L \times 2\gamma}$, which is defined as:

$$\underset{n_L \times 2\gamma}{\mathbf{P}} = \begin{bmatrix} \underset{n_L \times 1}{\mathbf{R}} & \Big| & \underset{n_L \times (\gamma-1)}{\mathbf{\Theta}} & \Big| & \underset{n_L \times \gamma}{\mathbf{N}} \end{bmatrix} = \begin{bmatrix} p_{(i,j)} \in \mathbb{R}^+ \end{bmatrix}_{\substack{i=1,2,\cdots,n_L \\ j=1,2,\cdots,2\gamma}} \quad (31)$$

Overall, within the expression of (31), the following essential deployment parameters are inscribed:

- number of deployment layers.
- width of each layer.
- number of sectors in each layer.
- extent of the angular boundary for each cluster.
- nodal scale randomly located in each sector.

At this level, we may harness the above descriptions by creating a generically flexible algorithm that enables controlled inhomogeneous random geometry. To emphasize, this method gives the necessary freedom to a cellular analyst or designer to selectively deploy random nodes in desired locations in order to form clusters. Once cluster-based random deployment is complete, the superposition principle can be applied to get the overall inhomogeneous spatial distribution of the cell.

All the required steps to accomplish the described ASD algorithm over a network model for the purpose of spatial inhomogeneity are provided in the pseudocode of Fig. 11. As evident by the nested for-loop, the algorithm is in part based on the foundation formulated for unbiased and exact random generation inside a flexibly versatile ring sector model derived and analyzed in Section 2.

On the whole, the conceptualized algorithm is a simple emulation tool useful for modeling a non-homogeneous network in instances when some elementary knowledge about a cell site is known or hypothesized. In fact, the treated inhomogeneous approach has the benefit of preserving full spatial randomness without relying on synthetic workarounds.

Now that we have the above pseudocode, it is noteworthy to determine by (32) the algorithm cost for executing this operation, where $n_S \in \mathbb{N}^*$ is the overall number of randomly deployed nodes within the network model.

$$O\left(\sum_{i=1}^{n_L}\sum_{j=1}^{n_{\text{sec}}^{(i)}} p_{(i,\ j+\gamma)}\right) = O\left(\sum_{i=1}^{n_L}\sum_{j=1}^{n_{\text{sec}}^{(i)}} n_{(i,j)}\right) = O\left(n_S\right) \quad (32)$$

### 4.2 Descriptive MC Deployment Examples

In Fig. 12, an example of a possible 3-layer network plan is shown. From the illustration, we identify that $n_{\text{sec-total}} = 6$ where the inner and outer layers have each a singular sector, and the middle layer is split into four clusters.

**Algorithm 1** - Emulating an Inhomogeneous Random Network - Controlled Footprint

1: **Require:** $n_L \in \mathbb{N}^+ \quad \left\{ n_{sec}^{(i)} \in \mathbb{N}^+ \right\} : i = 1, 2, \cdots, n_L \quad \mathbf{P} = \left[ p_{(i,j)} \in \mathbb{R}^+ \right]_{\substack{i=1,2,\cdots,n_L \\ j=1,2,\cdots,2\gamma}}$

2: **Compute:** $\gamma = \max\limits_{i=1,2,\cdots,n_L} \left\{ n_{sec}^{(i)} \right\}$

3: **Initialize :** $n_S = 0$

4: **for** $i = 1, 2, \cdots, n_L$ **do**

5:      **if** $\{i \neq 1\}$ **then** $L_1 := p_{(i-1,1)}$ **else** $L_1 := 0$ **end if**

6:      $L_2 := p_{(i,1)}$

7:      **for** $j = 1, 2, \cdots, n_{sec}^{(i)}$ **do**

8:          **if** $\{j \neq 1\}$ **then** $\alpha_1 := p_{(i,j)}$ **else** $\alpha_1 := 0$ **end if**

9:          **if** $\left\{ j \neq n_{sec}^{(i)} \right\}$ **then** $\alpha_2 := p_{(i,j+1)}$ **else** $\alpha_2 := 2\pi$ **end if**

10:          $n_0 := p_{(i,\, j+\gamma)}$

11:          **for** $m = 1, 2, \cdots, n_0$ **do**

12:              Generate two i.i.d. RVs: $\{\hat{u}_0, \hat{u}_1\} \sim \mathcal{U}(0,1)$

13:              Compute: $\hat{r}_m := \sqrt{L_1^2 + \hat{u}_0 \left( L_2^2 - L_1^2 \right)} \sim f_R(r)$

14:              Compute: $\hat{\theta}_m := \left\{ \alpha_1 + \hat{u}_1 (\alpha_2 - \alpha_1) \right\} \sim f_\theta(\theta)$

15:              Compute: $\left\{ \hat{x}_{n_S+m}; \hat{y}_{n_S+m} \right\} = \left\{ \hat{r}_m \cos\left(\hat{\theta}_m\right); \hat{r}_m \sin\left(\hat{\theta}_m\right) \right\} \sim f_{XY}(x, y)$

16:          **end for**

17:          $n_S := n_S + n_0$

18:      **end for**

19: **end for**

20: **Return:** $\{\hat{x}_t, \hat{y}_t\} : t = 1, 2, \cdots, n_S$

Fig. 11. Pseudocode for controlled heterogeneous deployment.

Also, in each of these zones, the amount of random nodes to be deployed is accordingly mapped. Although the derived inhomogeneous algorithm is scalable, in this deployment example we consider an $n_S = 3,300$ nodes.

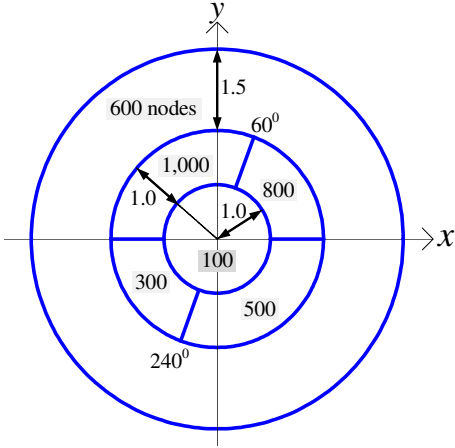

Fig. 12. Example of a 6-sector network footprint.

The network plan of Fig. 12 can equivalently be transformed into matrix format as follows:

$$\mathbf{P}_{3\times8} = \left[ \mathbf{R}_{3\times1} \,\middle|\, \mathbf{\Theta}_{3\times3} \,\middle|\, \mathbf{N}_{3\times4} \right] = \begin{bmatrix} 1.0 & 0 & 0 & 0 & 100 & 0 & 0 & 0 \\ 2.0 & \pi/3 & \pi & 4\pi/3 & 800 & 1{,}000 & 300 & 500 \\ 3.5 & 0 & 0 & 0 & 600 & 0 & 0 & 0 \end{bmatrix}$$

(33)

Fig. 13 shows the MC simulation result of the non-homogeneous network of (33). The generated structure is clearly a random network, i.e. this outcome is one of infinitely many random realizations of users' Euclidian geometry. This means that at every simulation run, the characterized network plan produces a unique inhomogeneous spatial emplacement. As for the corresponding

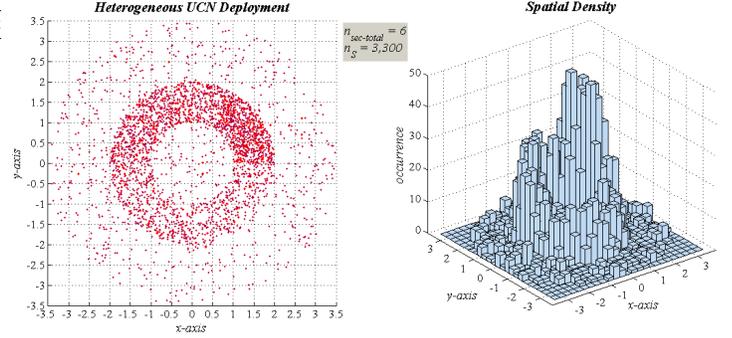

Fig. 13. Heterogeneous spatial deployment and density for a 6-sector cellular network example.

spatial density shown in the figure, it was estimated based on $25 \times 25$ grid.

To further display the conceptualized inhomogeneous algorithm of Fig. 11, we designed another cellular deployment with $n_S = 3,300$ nodes. This time however, the network is composed of 4-layers with $n_{\text{sec-total}} = 10$ sectors. The considered network footprint is depicted in Fig. 14, and its matrix equivalent is given in (34).

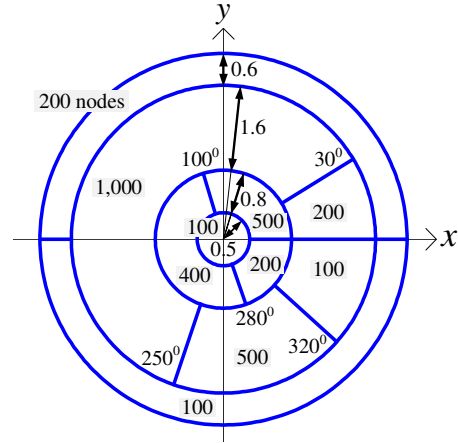

Fig. 14. Example of a 10-sector network footprint.

$$\mathbf{P}_{4\times8} = \begin{bmatrix} 0.5 & 0 & 0 & 0 & 100 & 0 & 0 & 0 \\ 1.3 & 5\pi/9 & 14\pi/9 & 0 & 500 & 400 & 200 & 0 \\ 2.9 & \pi/6 & 25\pi/18 & 16\pi/9 & 200 & 1{,}000 & 500 & 100 \\ 3.5 & \pi & 0 & 0 & 200 & 100 & 0 & 0 \end{bmatrix}$$

(34)

As a consequence of simulating this network, we obtain in Fig. 15 one of many possible random instances of the result.

When we compare the network plans of (34) with (33), we obviously notice that it has more entries, which in essence means that the associated spatial design is more elaborate than the previous one. In fact, the major elaboration of the network plan for a particular site is characterized by the $\mathbf{R}$ and $\mathbf{\Theta}$ components as a function of terrain features. Then, we could study and verify various QoS measures as the number of nodes in each sector is altered by a simple modification of the values in the $\mathbf{N}$ matrix.

Evidently, this approach may become handy when we couple to the spatial model a temporal element in order to further improve the modeling of users clustering tendency. And so, the nodal information of the network plan can accordingly be adjusted to reflect the changeable nature of users' geometry. In fact, these values could be conjectured based on plausible situations, or they could be compiled from simple statistical data gathering of a site as opposed to socially-intensive trend studies. Then, the ASD method can effortlessly be triggered to emulate a random spatial structure which can further assist in bridging the gap between reality and modeling.

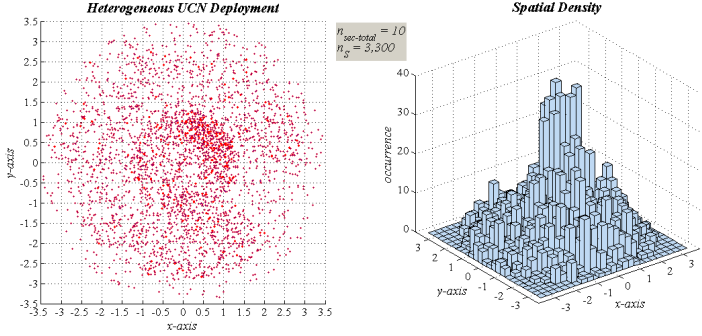

Fig. 15. Heterogeneous spatial deployment and density for a 10-sector cellular network example.

Before closing this section, we should remark that the model of the derived algorithm considers circular-based deployment strips. However, the various cluster sectors need not necessarily be of this form. Thus, the random deployment may not be appropriate at all times. Nonetheless, it could still be instrumental as an approximately more accurate deployment approach than currently available alternatives.

## 5  UNCONTROLLED AND AUTOMATIC ALGORITHM FOR RANDOM DEPLOYMENT

### 5.1 Formulation of the Network Model

From the above discussion, it should be obvious that the occurrence of clustering is inevitable in most real-life scenarios. As a result, the spatial distribution of nodes for a given deployment project will likely be non-homogeneous. For this reason, in the previous section we developed a practical spatial-level simulator tool for inhomogeneous random nodal deployment based on controlled network planning. While the approach is adequate, in particular cases, various modeling accommodations and extensions could be incorporated to this mechanism so that the emulation experience becomes more lucid for network designers. This endeavor will actually be the primary intention of the treatment that follows.

Specifically, we want to provide greater emulation leverage by conceptualizing another algorithm that can achieve heterogeneity with very limited planning information to the network subroutine. Thus, in contrast to the controlled approach of Section 4, the aim here is to construct an inhomogeneous random network in an uncontrolled or arbitrary manner. This could be done by redesigning the previous algorithm, in such a way that it maintains similar attributes, while requiring less input parameters in order to enable a simpler process for generating a heterogeneous spatial network.

From the ASD principle detailed in Section 3, we explained a strategy for inhomogeneity by ensuring different areal number densities in each of the deployment subregions. In fact, the density for the sectors is obtained by:

$$\forall A_i \in \mathbb{R}^*_+ : \exists n_i \in \mathbb{N}^* : \rho_i \triangleq n_i / A_i \quad i = 1, 2, \cdots, n_{\text{sec-total}} \quad (35)$$

From (35), we clearly notice that the densities can be unique in one of three possible ways:

1. vary $n_i$, and maintain $A_i$ fixed.
2. vary $A_i$, and maintain $n_i$ fixed.
3. vary simultaneously $n_i$ and $A_i$.

In deriving the desired inhomogeneous algorithm, we find that the second approach is more suitable; thus, (35) becomes: $\rho_i = n_0 / A_i \quad i = 1, 2, \cdots, n_{\text{sec-total}}$.

Next, in order to obtain different sub-regions, we will consider $n_L \in \mathbb{N}^*$ onion-like layers; therefore, for this layout $n_{\text{sec-total}} = n_L$. In fact, for the purpose of uncontrolled inhomogeneity, the number of layers will randomly be chosen from a predefined integer range. Therefore in [10], a generation technique for randomly producing discrete values from a generic uniform PMF, i.e.: $\Pr\{X = x\} = \mathcal{U}_D(n_1, n_2) \quad (n_1, n_2) \in \mathbb{Z}^2 : n_1 \le n_2$, was derived.

At present, we aim to randomly generate $n_L$ from a range delimited by $n_{L-\max}$, which essentially refers to the maximum arbitrary number of deployment layers possible for achieving spatial inhomogeneity. This value will actually be preset by the network designer at the start of the automatic emulation process. Thus, the number of layers at a simulation instance will be a RV specified by:

$$n_L \sim \mathcal{U}_D(2, n_{L-\max}) \qquad n_{L-\max} \in \mathbb{N}^* : n_{L-\max} > 1 \quad (36)$$

In (36), we notice that the sampling range begins at $n_1 = 2$ because from the ASD principle we at least need 2-layers for attaining inhomogeneity. In other words, if we would have started with $n_1 = 1$, and by discrete RNG $n_L$ is randomly set to this value, then we will simply obtain a homogeneous random network; this will actually be the antithesis to the wanted objective of spatial heterogeneity. Meanwhile, it is worth adding that in the rare but possible case where $n_{L-\max}$ is set to 2, then the number of layers will deterministically be assigned to this value.

Now that we have framed an approach for randomly obtaining the number of layers, the next step requires us to equally split the number of nodes among these subregions. By design, the overall amount of nodes $n_S \in \mathbb{N}^*$ planned for random deployment is supplied by the network architect. Since $n_S$ and $n_L$ need not necessarily be multiples of each other, then the number of nodes per layer must be arranged in a careful way. In particular, the amount of random nodes deployed in the innermost layer of an automatically emulated inhomogeneous network is designated by $n_{in} \in \mathbb{N}^*$. As for the outer layers, each of these sub-regions will contain $n_{out} \in \mathbb{N}^*$ nodes computed by: $\forall n_S \in \mathbb{N}^* : \exists n_L \in \mathbb{N}^* : n_{out} \triangleq \lfloor n_S / n_L \rfloor$. Knowing the volume of nodes in the outer layers, then it should be evident that the rest of the overall nodal quantity will constitute the amount of terminals in the innermost sub-region of the cell. Therefore, this measure can be calculated by:

$$n_{in} \triangleq n_S - (n_L - 1) \cdot n_{out} = n_S - (n_L - 1) \cdot \lfloor n_S / n_L \rfloor \quad (37)$$

So far, we have determined the number of layers and the amount of nodes in each sector. At present, we want to vary the areal size of each sub-region. This task can be done by randomly deciding on the geometrical position of the layers. That is, we want the width or thickness $\Delta_i \in \mathbb{R}_+^*$ of the various deployment layers to be different. In fact, this value corresponds to: $\Delta_i = r_i - r_{i-1} \quad i = 2, 3, \cdots, n_L$, such that $\Delta_1 = r_1$ is the radius measured from the origin of the Cartesian coordinate system to the first layer, and $r_i \in \mathbb{R}_+^*$ is the particular radius for all the other deployment layers. In this situation, the procedure to generate diverse widths of the deployment sub-regions can be realized by randomly producing radial values for the layers; this can be accomplished by: $r_i \sim \mathcal{U}_R(0, L) \quad i = 1, 2, \cdots, n_L - 1$. It should be clear that we only generate random radial values for the first $n_L - 1$ layers since $r_{n_L}$ will always be equal to the preassigned size of the cellular radius, namely $L \in \mathbb{R}_+^*$.

Following the generation of these radial distances, it becomes necessary to sort them in ascending order, i.e.:

$$\vec{r}_{sorted} = \text{sort}\left(\vec{r} \in \mathbb{R}_{+,*}^{n_L-1}\right) = \text{sort}\left([r_i]_{i=1,2,\cdots,n_L-1}\right) \quad (38)$$

There are many techniques available for implementing the sorting operator; some of the most notable among them are: quicksort, heapsort, and mergesort. Specifically, quicksort has been established as one of the fastest algorithms for ordering an array of numbers. Thus, MATLAB® uses this approach for its **sort** function.

Next, we will stochastically deploy in each of the formed random sized sub-regions the corresponding amount of nodes. Then, we superimpose these multi-density sectors together and look at the network as a holistic entity, which results into a heterogeneous outcome that has a random characteristic. With this conceptualization, we have progressively developed an automatic mechanism for randomly constructing the network plan so as to produce an inhomogeneous spatial structure. To be precise, the geometrical randomness is achieved due to the arbitrary nature of:

- the number of deployment layers: $n_L$
- the size of the layers: $\Delta_i$
- the position of nodes within each layer: $\{\hat{x}_i, \hat{y}_i\}$

Consequently, the amalgamation of the above factors will result into a heterogeneous random network. For the sake of completeness, these attributes are graphically depicted in the geometrical model of Fig. 16 used for automatically producing a random cellular network footprint.

Overall, the culmination of the above explanations and analysis enables us to derive the uncontrolled inhomogeneous algorithm of Fig. 17. From this algorithm, it can vividly be observed that a deployment designer will only require entering three essential inputs:

1. size of the cellular network: $L$
2. maximum number of deployment layers: $n_{L-\max}$
3. quantity of nodes to be deployed: $n_S$

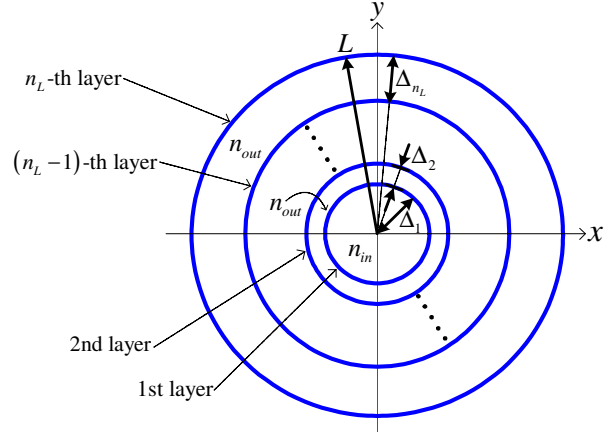

Fig. 16. Geometrical details for uncontrolled random deployment.

**Algorithm 2** - Emulating an Inhomogeneous Random Network - Uncontrolled Footprint

1:    Require: $n_S \in \mathbb{N}^+$   $L \in \mathbb{R}_+^*$   $\{n_{L-\max} \in \mathbb{N}^+ | n_{L-\max} > 1\}$
2:    Generate a RV: $\hat{u}_0 \sim \mathcal{U}(0,1)$
3:    Compute: $\hat{v}_0 = \{3/2 + \hat{u}_0 (n_{L-\max} - 1)\} \sim \mathcal{U}_V(3/2, n_{L-\max} + 1/2)$
4:    **for** $i = 2, 3, \cdots, n_{L-\max}$ **do**
5:      **if** $\{|\hat{v}_0 - i| \le 1/2\}$ **then**
6:        $n_L := i - \mathcal{U}_D (2, n_{L-\max})$ **break for**
7:      **end if**
8:    **end for**
9:    Compute: $n_{in} := n_S - (n_L - 1) \cdot \lfloor n_S / n_L \rfloor$   $n_{out} := \lfloor n_S / n_L \rfloor$
10:   **for** $j = 1, 2, \cdots, n_L - 1$ **do**
11:      Generate a RV: $\hat{u}_1 \sim \mathcal{U}(0,1)$
12:      Compute: $\hat{r}_j := \hat{u}_1 L \sim \mathcal{U}_R (0, L)$
13:   **end for**
14:   Sort in ascending order: $\{\hat{r}_j\} := \text{sort}(\hat{r}_j)$   $j = 1, 2, \cdots, n_L - 1$
15:   Initialize: $t = 0$
16:   **for** $j = 1, 2, \cdots, n_L$ **do**
17:      **if** $\{j \neq 1\}$ **then**
18:        $L_1 := \hat{r}_{j-1}$   $n_0 := n_{out}$
19:      **else**
20:        $L_1 := 0$   $n_0 := n_{in}$
21:      **end if**
22:      **if** $\{j \neq n_L\}$ **then** $L_2 := \hat{r}_j$ **else** $L_2 := L$ **end if**
23:      **for** $m = 1, 2, \cdots, n_0$ **do**
24:        Generate two i.i.d. RVs: $\{\hat{u}_2, \hat{u}_3\} \sim \mathcal{U}(0,1)$
25:        Compute: $\hat{r}_m := \sqrt{L_1^2 + \hat{u}_2 (L_2^2 - L_1^2)} \sim f_R(r)$
26:        Compute: $\hat{\theta}_m := 2\pi \hat{u}_3 \sim f_\theta(\theta)$
27:        Compute: $\{\hat{x}_{t+m}, \hat{y}_{t+m}\} = \{\hat{r}_m \cos(\hat{\theta}_m); \hat{r}_m \sin(\hat{\theta}_m)\} \sim f_{XY}(x, y)$
28:      **end for**
29:      $t := t + n_0$
30:   **end for**
31:   Return: $\{\hat{x}_i, \hat{y}_i\} : i = 1, 2, \cdots, n_S$

Fig. 17. Pseudocode for uncontrolled heterogeneous deployment.

Given that the formulated method only demands few entries, it then means that the heterogeneous algorithm of Fig. 17 will basically do most of the network decisions automatically in a stochastic way. In fact, when compared to the set of required parameters for the network footprint of the controlled deployment option detailed in Fig. 11, the discrepancy of the inputs among these inhomogeneous random network algorithms is considerable. In light of this reality, we can therefore remark that if less *a priori* information about the network project site is known or hypothesized, then the automatic inhomogeneous alternative for spatial emulation should be favored as opposed to the controlled algorithm.

In order to comprehend the time performance of the above algorithm, we find it necessary to evaluate in (39) its overall computational complexity. We should note that in this cost analysis, we considered $n_L = n_{L-\max}$ so as to reflect the worst computational scenario.

$$
\begin{aligned}
O\big(T_{\cos t}\big(L, n_{L-\max}, n_S\big)\big) &= O(1) + 2\cdot O\overbrace{\big(n_{L-\max}-1\big)}^{\triangleq m} \\
&+ O\big(\big(n_{L-\max}-1\big)\cdot \log_2\big(n_{L-\max}-1\big)\big) + O\big(n_S\big) \\
&\sim O\big(n_S\big) + O\big(m\big) + O\big(m\cdot \log_2 m\big) \\
&\sim O\big(n_S\big) + O\big(m\cdot \log_2 m\big) \\
&\sim O\big(n_S + n_{L-\max}\cdot \log_2 n_{L-\max}\big)
\end{aligned}
\tag{39}
$$

## 5.2 Descriptive MC Deployment Examples

At this level, it is interesting to highlight that the derived inhomogeneous algorithm of Fig. 17 can be used to automatically emulate a host of wireless network applications contained within a disk-shaped cellular lattice. In particular, it could be appropriate for stochastically mapping the spatial configuration of: wireless sensor networks (WSN), wireless mesh networks (WMN), or mobile networks. Indeed, each of these networks has a particular purpose and application focus. For instance, WSN is considered for low-power remote sensing; WMN is rather a multihop topology used for range extension or as a backup connectivity route; and cellular networks are aimed for ubiquitous long-range mobile communications [10]. In Table 2, the distinctive characteristics of these networks are accordingly outlined.

TABLE 2
On the Specifications of Different Wireless Networks

| network applications | communication range | P2MP network topology | examples of technologies | principal feature |
|---|---|---|---|---|
| WSN | meters | LR-WPAN | ZigBee, 6LoWPAN WirelessHART MiWi, LR-UWB | low-power sensor commuications |
| WMN | meters | WPAN WLAN | UWB, Bluetooth Wi-Fi | high-speed commuications |
| cellular network | kilometers | WMAN WWAN | MBWA, WiMAX LTE | long-range commuications |

Furthermore, these featured wireless networks are generally composed of variable-sized scale. For example, the volume of nodes in a WSN is for the most part way layer than a WMN because point sensors are typically cheaper to fabricate. Specifically, the scale of a WSN is somewhere in the order of hundreds up to thousands, and could in extreme cases reach millions of nodes [18], [19]. Irrespective of the quantity of nodes to be deployed, the formulated inhomogeneous random network algorithm is scalable for the emulation scenario under study. Given that the algorithm supports diverse spatial geometries, we will therefore demonstrate the scalability aspect by generating various heterogeneous random realizations of the network model.

In Fig. 18, we show four random instances of a small-scale heterogeneous deployment. Within these results, in addition to the actual spatial deployment, the network plan is separately graphed so as to emphasize its arbitrary nature. In other words, the planning of the network, which is assembled by the number of layer, the deployment size, and the geometry of nodes, is randomly obtained in an automatic way at every simulation run.

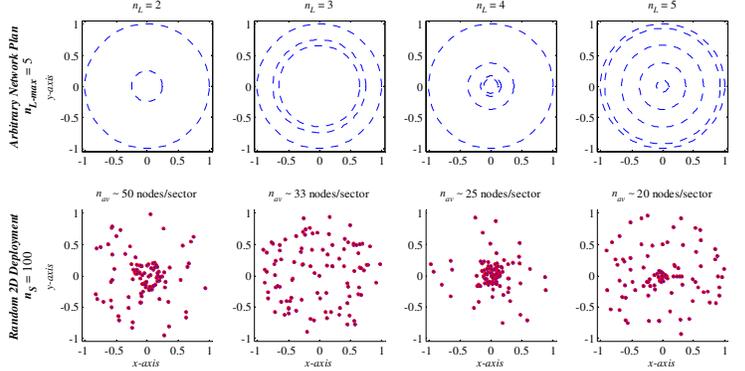

Fig. 18. Random instances of small-scale deployment.

Similarly, in Fig. 19, we demonstrate another set of examples for a medium-scale network. Again, each run of the simulation produces a unique inhomogeneous random spatial realization.

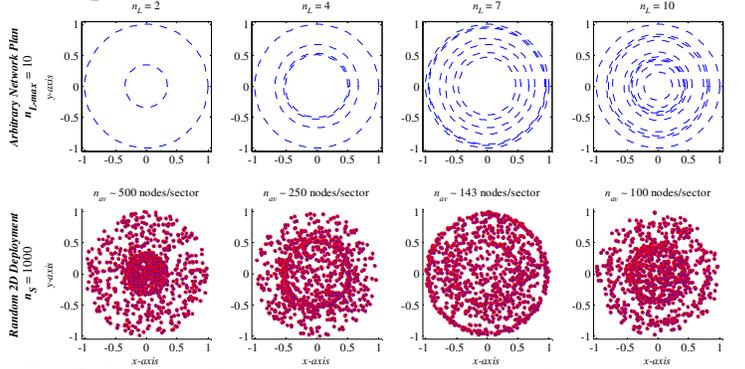

Fig. 19. Random instances of medium-scale deployment.

As a final representative example, a large-scale model for the network is emulated in Fig. 20. Because the scale is relatively elevated when compared to the other two cases, as illustrated by this MC simulation, the geometrical resolution of each node is reduced.

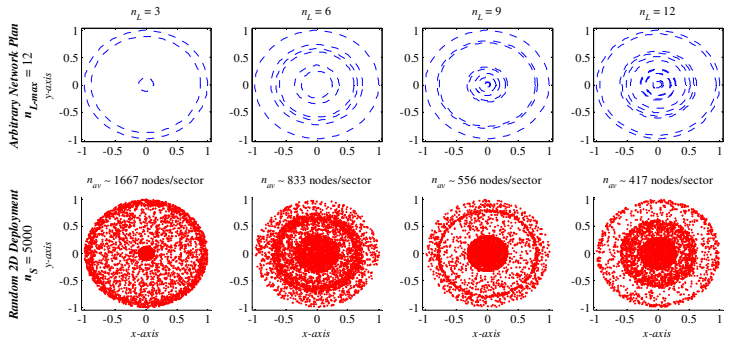

Fig. 20. Random instances of large-scale deployment.

It is valuable to note that in emulating an uncontrolled inhomogeneous architecture, there is no particular interconnection, linear or otherwise, between the inputted maximum number of deployment layers and the network scale; unless such relation is intentionally assumed by the designer. From Table 3, which essentially summarizes the considered inputs to the network instances generated in Figs. 18, 19 and 20, we in fact hypothesized such correlation between the supplied variables. That is, for the three network cases, as $n_{L-max}$ increased, by design the network scale $n_s$ also rose.

TABLE 3
SIMULATION INPUTS USED FOR AUTOMATIC DEPLOYMENT

| network models | $L$ (unit) | $n_{L-max}$ (no.) | $n_s$ (no.) |
|---|---|---|---|
| small-scale | 1 | 5 | 100 |
| medium-scale | 1 | 10 | 1,000 |
| large-scale | 1 | 12 | 5,000 |

Although there is no explicit association among $n_{L-max}$ and $n_s$; yet on the other hand, we notice a symbiotic relationship between $n_{L-max}$ and the degree of network inhomogeneity that requires some carefully calculated scrutiny. To be precise, the direct interdependence with $n_{L-max}$ and the random number of layers is indicated in (36). And, as $n_l$ increases, the amount of random sized subregions will straightforwardly augment. Assuming that the number of nodes per sector remains steady; consequently, the quantity of layers with unique densities will also rise. Thus, for all practical purposes, it is projected that $n_{L-max}$ affects the inhomogeneity of a random network constellation. However, more research wok is still required in order to analytically describe and quantify the extent of the heterogeneity.

## 6 SYNOPSIS OF THE SIMULATOR MODELS

Overall, the characteristics and benefits of the simple ASD modeling algorithm for spatial inhomogeneity can be summarized as follows:

- The spatial deployment is more realistic than other alternatives; thus resulting a more reflective tool which can be informative for network planning and service arrangements.
- It is a cost and time efficient approach for heterogeneous deployment because the modeling is only based on limited *a priori* network planning inputs that can be obtained via simple pattern analysis or feasible conjectures.
- The deployment method is flexible, in the sense that the precision and complexity of the inhomogeneous random graph to be generated is left at the discretion of the network designer. In other words, if more network planning inputs are provided, then the spatial accuracy of the corresponding emulation will further be enhanced.
- The ASD deployment supports different random network models in an entirely generic manner through the use of variable entries for:

  - o Controlled Algorithm: the number of deployment layers, the width of each layer, the number of sectors in each layer, the extent of the angular boundary for each cluster, and the nodal scale per sector.
  - o Uncontrolled Algorithm: the cellular size, the maximum arbitrary number of deployment layers, and the overall nodal scale.

- The inhomogeneous simulator algorithms are structured in a systematic and modular way, which makes them reasonably practical for implementation using programming packages.
- The algorithms are also coherently formulated, thus parameter modifications can be configured in a straightforward manner.
- Testing and repeatability of the random experimentation is easily possible for execution.
- On the whole, the simple inhomogeneous spatial models can be practical during planning in order to inquire and evaluate the impact of different spatial deployments on QoS metrics, so as to enhance and optimize the network performance by strategically designing the architecture.

## 7 CONCLUSION

The importance of all variations of wireless communications, and in particular cellular technologies, are still and even more significant as we move toward newer network generations. Therefore, analysis and planning of such systems through time- and cost-efficient simulations are vital. As a result, the central focus of this paper was based on the random emulation of terminals spatial position.

In fact, we remarked that typical spatial distribution densities, though practical to some degree, have their own limitations. Therefore, an inhomogeneous deployment algorithm based on the superposition principle of targeted spatial distribution was proposed. This conceptualized heterogeneous networking approach, which we refer to as ASD, is certainly more manageable because it breaks down a fairly complicated task of finding the wholesome density of users' spatial pattern in a vast terrain to that of smaller regions. Then, the principle of superposition can be applied to merge the spatial clusters together, and hence establish the entire random mobile distribution of the cell in order to investigate various network-based integrity measures. Overall, this controlled spatial emulation algorithm is a coherent, easily configured tool, with greater emulation flexibility, useful for effectively modeling and attaining a heterogeneous random arrangement.

In addition to the above controlled deployment algorithm, we derived an automatic ASD emulator to arbitrarily simulate an inhomogeneous wireless network. This uncontrolled heterogeneous spatial generator method is practical when insufficient or no specific information about a network site is known or asserted; i.e., the designer is not entirely aware of the actual deployment environment. The key advantage of this tool is that it can randomly construct a unique heterogeneous geometry suitable for small, medium or large scale networks while necessitating very few input parameters.

In general, these simple inhomogeneous simulator models can be used for studying a host of factors that affects the link-layer of the network, among others: channel losses, interference, and resource consumption.

**Mouhamed Abdulla** received, respectively in 2002, 2005, and 2012, a B.Eng. (with Distinction) in electrical engineering, an M.Eng. in aerospace engineering, and a Ph.D. in telecommunications all at Concordia University in Montréal, Québec, Canada. He is currently a systems engineering researcher with the Department of Electrical and Computer Engineering at Concordia University. Moreover, for nearly 7 years since 2003, he worked at IBM Canada Ltd. as a senior technical specialist. Dr. Abdulla holds several awards and honors from academia, government, and industry; among them the Golden Key Outstanding Scholastic Achievement and Excellence Award (2009), and the IBM Innovation Award (2007). He is professionally affiliated with IEEE, IEEE ComSoc, IEEE GOLD, ACM, AIAA, and OIQ. Currently, he is a member of the executive committee for IEEE Montréal section, and serves as its Secretary. In addition, he is also the Secretary of IEEE ComSoc and ITSoc societies. Furthermore, he is an associate editor of the *IEEE Technology News Publication*, *IEEE AURUM Newsletter*; and editor of *Journal of Next Generation Information Technology*, and *Advances in Network and Communications Journal*. Also, he regularly serves as a referee for a number of journal publications such as: IEEE, Springer, and EURASIP; and contributes as an examination item writer for the prominent *IEEE/IEEE ComSoc WCET® Certification Program*. Besides, he constantly serves as a technical program committee (TPC) member for several IEEE international conferences. Presently, his research is focused on advancing the fundamentals and characteristics of wireless random networks. Moreover, he has a particular interest in philosophical factors and social aspects related to engineering education, and research innovation. Since 2011, his biography is listed in the distinguished Marquis Who's Who in the World publication.

**Yousef R. Shayan** received his Ph.D. degree in electrical engineering from Concordia University in 1990. Since 1988, he has worked in several wireless communication companies at different capacities. He has been in the R&D departments of SR Telecom, Spar Aerospace, Harris, and BroadTel Communications, a company he co-founded. In 2001, he joined the department of Electrical and Computer Engineering of Concordia University as associate professor. Since then he has been Graduate Program Director, Associate Chair, and Department Chair. Prof. Shayan is founder of the Wireless Design Laboratory, which was established in 2006 based on a major CFI Grant. This lab has state-of-the art equipment which is used for development of wireless systems. In June 2008, he was promoted to the rank of professor, and was also recipient of the Teaching Excellence Award for academic year 2007-2008 awarded by Faculty of Engineering and Computer Science. Since 1985, he has contributed as a technical program committee member for a number of major IEEE conferences. In 2004, he was elected as IEEE Chair for the communications chapter of Montréal, and in 2007 and 2008 was the Treasurer. Lately, in 2010, he was the Chair of the 23rd IEEE CCECE Communications and Networking Symposium; and in 2012, he was the treasurer of this flagship conference. His fields of interests include: wireless networking, error control coding, and modulation techniques.